# Rigorous electron affinity determination of complex heavy atoms and fullerene molecules

Zineb Felfli and Alfred Z. Msezane

Department of Physics and Center for Theoretical Studies of Physical Systems, Clark Atlanta University, Atlanta, Georgia 30314, USA

**Abstract**

Regge-pole calculated low-energy electron elastic total cross sections (TCSs) for complex heavy atoms and fullerene molecules are characterized generally by ground, metastable, and excited negative-ion formation, shape resonances and Ramsauer-Townsend minima. Here the extracted anionic binding energies (BEs) from the TCSs of various atoms and fullerenes are used to highlight the ambiguous meaning of some current electron affinities (EAs) of heavy complex atomic systems. The crucial question is: does the EA correspond to the BE of the attached electron in the ground or excited state of the formed anion during the collision?

## 1. Introduction

This paper is motivated by the important question: *Does the* electron affinity (EA) of complex many-electron heavy atoms and fullerene molecules *correspond to the binding energy (BE) of the attached electron in the ground, metastable or excited state of the formed negative ion during the collision?* Clarification of this issue should lead to unambiguous and definitive meaning of the EA of complex heavy systems.

In this paper we discuss the meaning of measured EAs of complex heavy systems (atoms and fullerene molecules) within the context of two prevailing viewpoints: 1) The first considers the EA to correspond to the electron BE in the ground state of the formed negative ion during collision; it is exemplified by the measured EAs of Au, Pt and At atoms and the fullerene molecules from $C_{20}$ through $C_{92}$; and 2) The second view identifies the measured EA with the BE of electron attachment in an excited state of the formed anion. The measured EAs of Ti, Hf, lanthanide and actinide atoms provide representative examples of this viewpoint. Here the extracted anionic BEs from the Regge-pole calculated low-energy electron elastic total cross sections (TCSs) for Au, Pt, At, Ti, Hf, lanthanide (Nd, Eu an Tm) and actinide (Th, U, Am and Lr) atoms and the fullerene molecules ($C_{28}$, $C_{60}$ and $C_{70}$) are contrasted with the measured and/or calculated EAs to highlight ambiguous and confusing meaning of the EA.

Low-energy electron collisions, resulting in negative ion formation, provide a special insight into quantum dynamics [1]. Consequently, needed is the careful determination of EAs, the Ramsauer-Townsend (R-T) effect, important for understanding sympathetic cooling and production of cold molecules using natural fermions, the Wigner threshold law, essential in high precision measurements of BEs of valence electrons and shape resonances (SRs). Accurate and reliable atomic and molecular affinities are essential for understanding chemical reactions involving negative ions [2]. Unfortunately, for most of the lanthanide atoms, producing sufficient anions that can be used in photodetachment experiments is challenging [3]. Due to their radioactive nature the actinide atoms are difficult to handle experimentally. Thus the great need for reliable theoretical EAs to guide measurements.

The EAs of atomic Au, Pt, and At have been measured [4-9], including those of the $C_{60}$ and $C_{70}$ fullerene molecules [10-14]. For the highly radioactive At atom various sophisticated theoretical calculations, including the Multiconfiguration Dirac Hartree-Fock (MCDHF) value [15] agree excellently with the measured EA [9]. Reference [9] employed Coupled-Cluster method, while Ref.[15] used MCDHF method. Furthermore, in [15] an extensive comparison among various sophisticated theoretical EAs has been carried out. For all these atoms the measured EAs matched excellently the Regge-pole calculated BEs of the anionic ground states of the formed negative ions during the collisions, see Table 1 for comparisons. Also, the measured EAs of the fullerene molecules $C_{20}$ through $C_{92}$ agree excellently with the Regge-pole calculated anionic ground states BEs [16, 17]. This gives great credence to our interpretation of the EAs of these heavy complex systems, viz. as corresponding to the ground states BEs of the formed negative ions during the collisions.

Recently, the EAs of Th[18] and U[19, 20] were measured as well. The authors remarked that the EAs of both Th and U corresponded to the BEs of the weakly bound electron to the neutral atoms. For the Ti atom two measurements obtained the EAs as 0.377 eV[21] and 0.075 eV[22]. The former value is close to various theoretical calculations [23, 24]**,** including the Regge-pole calculated BE of the second excited state, namely 0.281 eV[25]. However, the value of 0.075 eV[22] is close to the Regge-pole BE of the highest excited state of the formed Ti⁻ anion, 0.0664 eV; its ground state BE is 2.42 eV [25]. The measured EA of Hf is 0.178 eV[26]. It is close to the Regge pole SR at 0.232 eV, the RCI EA of 0.114 eV[27] and the Regge-pole second excited state anionic BE of 0.113 eV[28]. The Hf highest excited state BE is at 0.017 eV[29]. Indeed, here we are faced with the problem of interpretation of what is meant by the EA.

For the lanthanide atoms problems regarding what is meant by the EA have already been discussed[29, 30]. Briefly, for the Nd atom, there are two measured EA values, viz. 1.916 eV [31] and 0.0975 eV [32]. The value of [31] is close to the Regge-pole ground state anionic BE value of 1.88 eV [33], while the EA of [32] is close to the RCI EA [34] and the Regge –pole anionic BE of the highest excited state [33]. Similarly the measured EAs for the Eu atom are 0.116 eV[3] and 1.053 eV[35]. The former value agrees excellently with the Regge-pole BE of the highest excited state, viz. 0.116 eV[33], see also Fig. 3 here and with the RCI EA of 0.117 eV[36]. The EA of [35] agrees very well with the Regge-pole metastable anionic BE value of 1.08 eV[33]. For the Tm atom the measured EA[37] is close to the Regge-pole metastable BE(1.02 eV) [33]. Clearly, the results here demonstrate the need for an unambiguous meaning of the EA of the complex heavy atoms.

## 2. Method of calculation

In this paper we have used the rigorous Regge pole method to calculate elastic total cross sections (TCSs). Regge poles, singularities of the S-matrix, rigorously define resonances [38,39] and in the physical sheets of the complex plane they correspond to bound states [40]. In [41] it was confirmed that Regge poles formed during low-energy electron elastic scattering become stable bound states. In the Regge pole, also known as the complex angular momentum (CAM), method the important and revealing energy-dependent Regge trajectories are also calculated. Their effective use in low-energy electron scattering has been demonstrated in for example [33,42].

The near-threshold electron–atom/fullerene collision TCS resulting in negative-ion formation as resonances is calculated using the Mulholland formula [43]. In the form below, the TCS fully embeds the essential electron-electron correlation effects [44, 45] (atomic units are used throughout):

$$\sigma_{tot}(E) = 4\pi k^{-2} \int_0^\infty \text{Re}[1 - S(\lambda)]\lambda d\lambda$$
$$- 8\pi^2 k^{-2} \sum_n \text{Im} \frac{\lambda_n \rho_n}{1 + \exp(-2\pi i \lambda_n)} + I(E) \quad (1)$$

In Eq. (1) $S(\lambda)$ is the S-matrix, $k = \sqrt{2mE}$, $m$ being the mass and $E$ the impact energy, $\rho_n$ is the residue of the S-matrix at the $n^{th}$ pole, $\lambda_n$ and $I(E)$ contains the contributions from the integrals along the imaginary $\lambda$-axis ($\lambda$ is the complex angular momentum); its contribution has been demonstrated to be negligible [33].

As in [46] here we consider the incident electron to interact with the complex heavy system without consideration of the complicated details of the electronic structure of the system itself. Therefore, within the Thomas-Fermi theory, Felfli et al [47] generated the robust Avdonina-Belov-Felfli (ABF) potential which embeds the vital core-polarization interaction

$$U(r) = -\frac{Z}{r(1+\alpha Z^{1/3}r)(1+\beta Z^{2/3}r^2)} \quad (2)$$

In Eq. (2) $Z$ is the nuclear charge, $\alpha$ and $\beta$ are variation parameters. For small $r$, the potential describes Coulomb attraction between an electron and a nucleus, $U(r) \sim -Z/r$, while at large distances it has the appropriate asymptotic behavior, viz. $\sim -1/(\alpha\beta r^4)$ and accounts properly for the polarization interaction at low energies. For an electron, the source of the bound states giving rise to Regge trajectories is the attractive Coulomb well it experiences near the nucleus. The addition of the centrifugal term to the well 'squeezes' these states into the continuum [45].

The strength of this extensively studied potential [48] lies in that it has five turning points and four poles connected by four cuts in the complex plane. The presence of the powers of $Z$ as coefficients of $r$ and $r^2$ in Eq. (2) ensures that spherical and non-spherical atoms and fullerenes are correctly treated. Also appropriately treated are small and large systems. The effective potential $V(r) = U(r) + \lambda(\lambda+1)/2r^2$ is considered here as a continuous function of the variables $r$ and complex $\lambda$. The details of the numerical evaluations of the TCSs have been described in [45] and references therein.

## 3. Results

Figure 1, taken from Ref. [49] presents TCSs for atomic Au and fullerene molecule $C_{60}$. They typify the TCSs of complex heavy atoms and fullerene molecules, respectively. Importantly, they are characterized by dramatically sharp resonances representing negative-ion formation in ground, metastable and excited anionic states, R-T minima and SRs. In both Figs. the red curves represent ground states electron TCSs while the green curves denote excited states TCSs. Here the ground states anionic BEs in both Au and $C_{60}$ appearing at the absolute R-T minima matched excellently the measured EAs, see Table 1 for comparisons with various measurements. In both systems, the ground states anionic BEs determine their EAs and not the excited anionic BEs (green curves). The data in Table 1 for Pt and At atoms as well as for the fullerene molecules were extracted from similar curves as in Fig. 1. Notably, the TCSs for the atoms and fullerene molecules become more complicated as the systems considered become larger as exemplified by the actinide atoms in Fig. 4

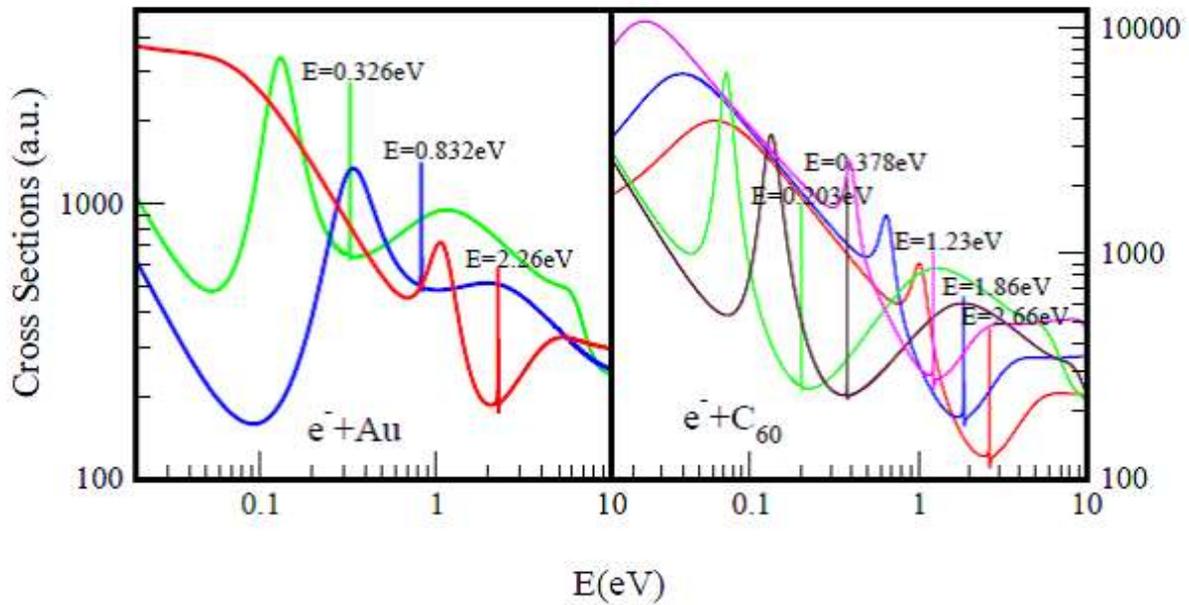

**Figure 1:** Total cross sections (a.u.) for electron elastic scattering from atomic Au (left panel) and the fullerene molecule $C_{60}$ (right panel) are contrasted. For atomic Au the red, blue and green curves represent TCSs for the ground, metastable and excited states, respectively. For the $C_{60}$ fullerene the red, blue and pink curves represent TCSs for the ground and the metastable states, respectively while the brown and green curves denote TCSs for the excited states. The dramatically sharp resonances in both figures correspond to the $Au^-$ and $C_{60}^-$ negative-ions formation.

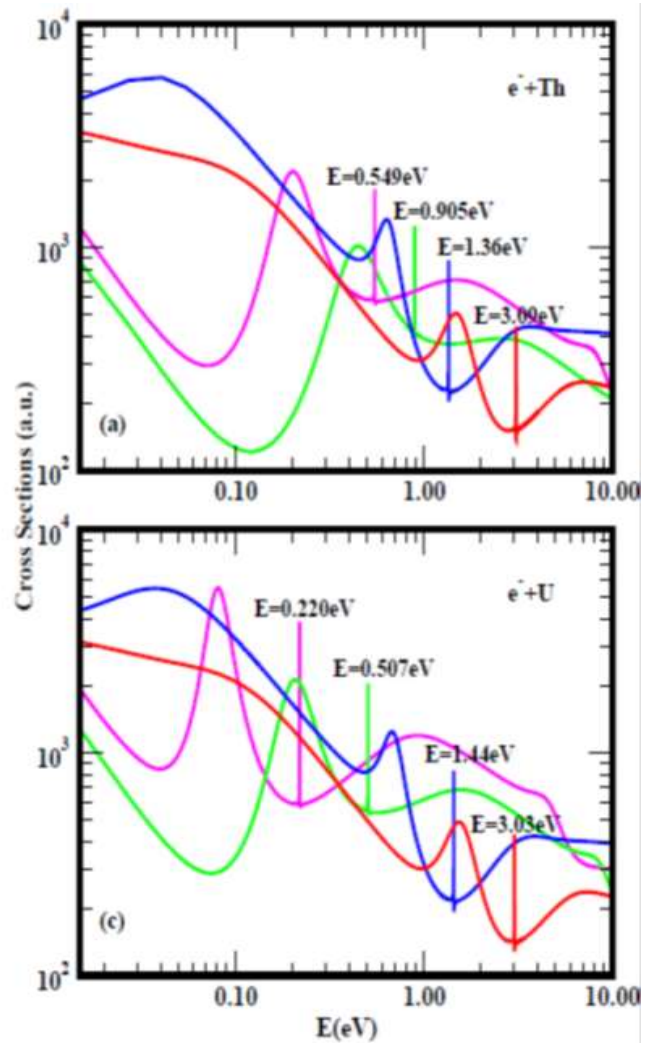

Figure 2, taken from Ref. [49] with a slight modification presents the TCSs for atomic Th and U. They typify the TCSs of the smaller complex heavy actinide atoms. Importantly, they are characterized by dramatically sharp resonances representing negative-ion formation in ground, metastable and excited anionic states, R-T minima and SRs. In both Figs. the red curves represent electron attachment in the ground states while the pink curves denote excited states curves. For Th, Fig. 2 top the measured and calculated EA values are 0.608 eV and 0.599 eV[18], respectively. These values are close to the Regge-pole anionic BE of the second excited state, pink curve (0.549 eV). Close to this value there is a SR at 0.61 eV defined by the blue curve. Not shown is the highest excited state curve with anionic BE value of 0.149 eV.

**Figure 2:** Total cross sections (TCSs) for atomic Th (top) and U(bottom) curves. The relevant curves in both TCSs are the ground states (red curves) and the excited states (pink curves). The dramatically sharp peaks in both Figs. with attendant BEs represent electron attachment.

The EAs of U have been measured very recently to be 0.315 eV [19] and 0.309 eV [20] as well as calculated to be 0.232 eV [20]. These values are close to the Regge pole anionic BE value of 0.220 eV for the highest excited state, see Table 1 for additional comparisons.

Here we do not understand the inconsistency in the meaning of the EAs in Figures 1 and 2, namely corresponding to the BEs of electron attachment in ground and excited anionic states, respectively.

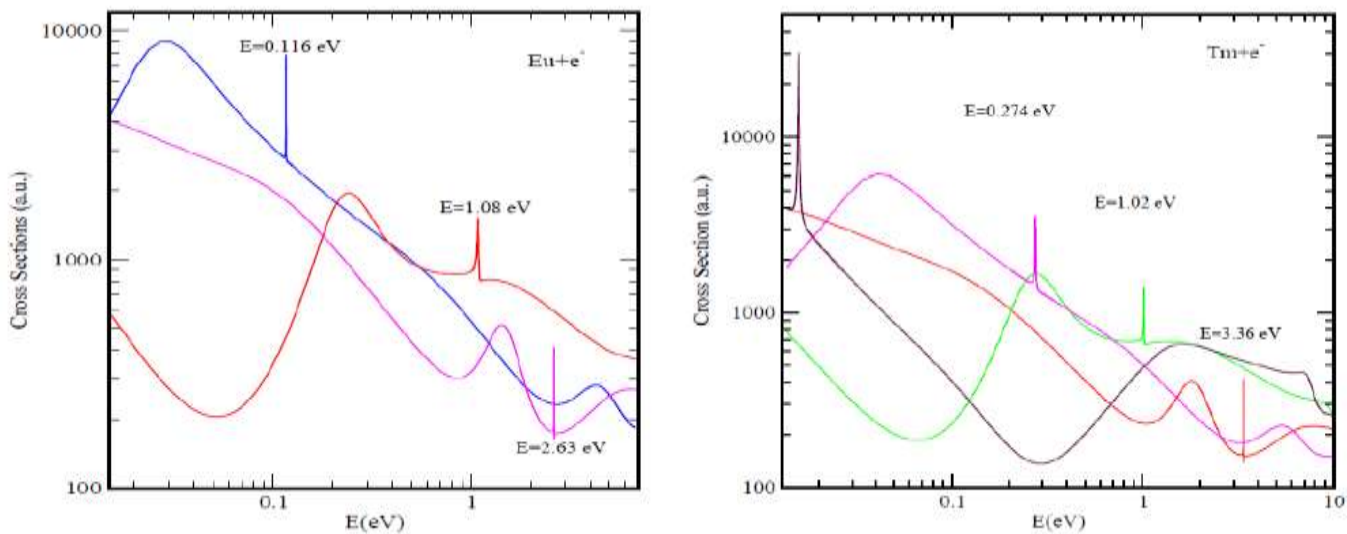

**Figure 3:** Total cross sections (TCSs) for the lanthanide atoms Eu (left panel) and Tm (right panel) curves. The relevant curves here in both TCSs are the ground states (pink and red curves, respectively) and the excited states (blue and pink curves, respectively). The dramatically sharp peaks in both Figs. with attendant BEs represent electron attachment. The very sharp peak in the near threshold TCSs for Tm has a BE value of 0.016 eV.

Figure 3, taken from [30], with the second Figure recalculated, presents the TCSs for the lanthanide atoms Eu (left panel) and Tm (right panel). The relevant curves here in both TCSs are the ground states (pink and red curves, respectively) and the excited states (blue and pink curves, respectively). The dramatically sharp peaks in both Figs. with attendant BEs represent electron attachment. For Eu we focus on the ground state, pink curve with a BE value of 2.63 eV and the blue curve with the BE of 0.116 eV, corresponding to an excited state TCS. The measured EA (0.116 eV)[3] is in outstanding agreement with the excited state BE value above and the RCI calculated EA (0.117 eV)[36], see Table 1. The metastable BE value of 1.08 eV, red curve agrees excellently with the measured EA (1.053 eV)[35]. This clearly demonstrates the ambiguous and confusing meaning of the measured EA of Eu by Refs. [3] and [35]. Does the EA of Eu correspond to the BE of electron attachment in the metastable state or the excited state of the formed anion during the collision?

Similarly with the case of the Tm atom; the Regge pole calculated ground and excited states BEs are respectively 3.36 eV and 0.274 eV. The measured EA of Tm is 1.029 eV[37] and agrees excellently with the Regge pole calculated metastable BE value of 1.02 eV, green curve in the figure. Accordingly, here the meaning of the measured EA of Tm corresponds to the BE of the metastable state. Indeed, in both Eu and Tm atoms the meaning of the measured EAs is both ambiguous and confusing as well.

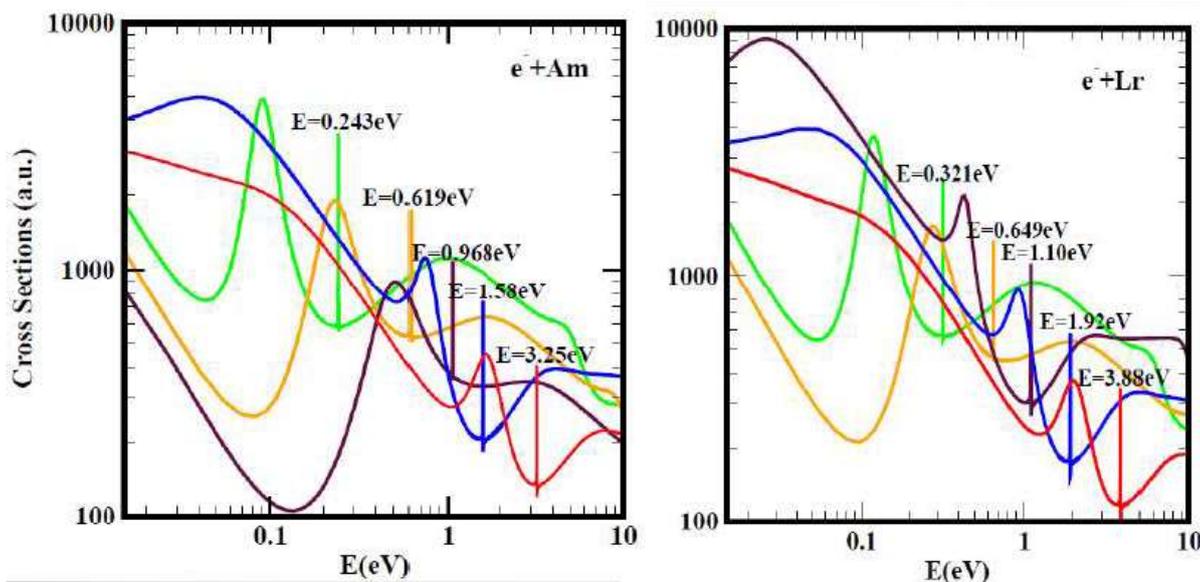

**Figure 4:** Total cross sections (TCSs) for the large actinide atoms Am (left panel) and Lr (right panel). The relevant curves here in both TCSs are the ground states (red curves) and the excited states (green curves). The dramatically sharp peaks in both Figs. with attendant BEs represent electron attachment. The brown curves in the TCSs for Am and Lr manifest the importance of the size and orbital collapse effects [50].

The TCSs in Fig. 4, taken from [50] are for the large actinide atoms Am (left panel) and Lr (right panel). The relevant curves here in both TCSs are the ground states (red curves) and the excited states (green curves). The dramatically sharp peaks in both Figs. with attendant BEs represent electron attachment. The brown curves in the TCSs for both Am and Lr manifest the importance of the size and orbital collapse effects[50]. There are no measured EAs for both atoms to compare our BEs with. However, theoretical EAs are available [36, 51, 52, 53] to compare with, see Table 1. For atomic Lr the EA values of 0.310 eV[51] and 0.295 eV[36] are close to the Regge-pole BE of the highest excited state, namely 0.321 eV. Notably, there are SRs close to the 0.321 eV BE. These values should guide the reliable measurement of the EA of Lr. Also included in Table 1 for comparisons are the measured EAs for the small $C_{28}$ [54, 55], the theoretical BEs for Au [56] and Pt[56] as well as the calculated EAs of At[57], $C_{28}$ [58], $C_{60}$ [59] and $C_{70}$ [60].

## 4. Conclusion

To prevent the proliferation of ambiguous and confusing meaning of the EAs of complex heavy atoms now populating the published literature, a definitive answer is required to this simple question: Does the EA of complex heavy atoms and fullerenes correspond to the BE of the attached electron in the ground or excited state of the formed negative ion during the collision?

**Table 1:** Negative-ion binding energies (BEs) and ground states Ramsauer-Townsend (R-T) minima, all in eV extracted from TCSs of the atoms Au, Pt, At, Ti, Hf, Th, U, Am, Lr, Nd, Eu, and Tm and the fullerene molecules $C_{28}$, $C_{60}$ and $C_{70}$. They are compared with the measured electron affinities (EAs) in eV. GRS, MS-$n$ and EXT-$n$ ($n=1, 2$) refer respectively to ground, metastable and excited states. Experimental, EXPT and theoretical, Theory EAs including RCI [36] and GW[53] are also included.

| System Z | BEs GRS | BEs MS-1 | BEs MS-2 | EAs EXPT | BEs EXT-1 | BEs EXT-2 | R-T GRS | BEs/EAs Theory | EAs [36] | EAs [53] |
|---|---|---|---|---|---|---|---|---|---|---|
| Au 79 | 2.26 | 0.832 | - | 2.309[4] 2.301[5] 2.306[6] | 0.326 | - | 2.24 | 2.262[56] | - | - |
| Pt 78 | 2.16 | 1.197 | - | 2.128[4] 2.125[7] 2.123[8] | 0.136 | - | 2.15 | 2.163[56] | - | - |
| At 85 | 2.41 | 0.918 | - | 2.416[9] | 0.115 | 0.292 | 2.43 | 2.38[15] 2.42[57] | - | - |
| Ti 81 | 2.42 | - | - | 0.377[21] 0.075[22] | 0.066 | 0.281 | 2.40 | 0.27 [23] 0.291 [24] | - | - |
| Hf 72 | 1.68 | 0.525 | - | 0.178[26] | 0.017 | 0.113 | 1.67 | 0.114[27] 0.113[28] | - | - |
| $C_{28}$ | 3.10 | 1.80 | 0.305 | 2.80[54] 3.00[55] | - | - | 2.98 | 3.39[58] | - | - |
| $C_{60}$ | 2.66 | 1.86 | 1.23 | 2.684[10] 2.666[11] 2.689[12] | 0.203 | 0.378 | 2.67 | 2.663[17] 2.63[59] | - | - |
| $C_{70}$ | 2.70 | 1.77 | 1.27 | 2.676[11] 2.72[13] 2.74[14] | 0.230 | 0.384 | 2.72 | 3.35[60] 2.83[60] | - | - |
| Th 90 | 3.09 | 1.36 | 0.905 | 0.608[18] | 0.149 | 0.549 | 3.10 | 0.599[18] | 0.368 | 1.17 |
| U 92 | 3.03 | 1.44 | 1.22 | 0.315[19] 0.309[20] | 0.220 | 0.507 | 3.01 | 0.175[36] 0.232 [20] | 0.373 | 0.53 |
| Am 95 | 3.25 | 1.58 | 0.968 | N/A | 0.243 | 0.619 | 3.27 | - | 0.076 | 0.142 |
| Lr 103 | 3.88 | 1.92 | 1.10 | N/A | 0.321 | 0.649 | 3.90 | 0.310[51] 0.476[52] | 0.465 0.295 | -0.212 -0.313 |
| Nd 60 | 1.88 | 1.73 | 0.997 | >1.916[31] 0.0975[32] | 0.162 | 0.505 | 1.90 | 0.162[33] 0.167[34] | - | - |
| Eu 63 | 2.63 | 1.08 | - | 0.116[3] 1.053[35] | 0.116 | 0.619 | 2.62 | 0.116[33] 0.117[36] | - | - |
| Tm 69 | 3.36 | 1.02 | - | 1.029[37] | 0.016 | 0.274 | 3.38 | - | - | - |

**Acknowledgment**
Research was supported by the U.S. DOE, Division of Chemical Sciences, Geosciences and Biosciences, Office of Basic Energy Sciences, Office of Energy Research, Grant: DE-FG02-97ER14743.